# Transitions in ZnS and CdSe quantum dots and wave-function symmetry

B. Zorman
*The Department of Physics, Columbia University, New York, New York 10027
and The Department of Chemistry, Columbia University, New York, New York 10027-6948*

R. A. Friesner
*The Department of Chemistry, Columbia University, New York, New York 10027-6948*

Excitation energies for wurtzite spherical ZnS and CdSe quantum dots in the range of 40–4000 atoms were calculated using empirical pseudopotentials and a real-space basis. The energies are compared to experiments and other pseudopotential models. For ZnS quantum dots, squared transition dipole sums were computed efficiently, without the need for full wave functions of the excited states; and some transition dipole calculations include the effects of an approximate electron-hole Coulomb potential. Squared transition dipole sums from the highest energy linear dipole like valence states to the lowest excited state were computed as a function of dot size. The model predicts that the per atom dipole transition sum decreases with quantum dot size for those transitions. The mixing of even and odd angular components and charge asymmetry of the wave functions affect the dipole transition strengths. The total oscillator strength for the lowest energy transition region increases with size at small radii, resembling the pattern recently observed experimentally for CdSe quantum dots. We examined the role of wave-function angular momentum for transitions to conduction band surface states. © *2003 American Institute of Physics.* [DOI: 10.1063/1.1557178]

## I. INTRODUCTION

Research on the optical excitation of electrons in semiconductor quantum dots has often been sparked by potential applications of these structures. CdSe nanostructures can appear in light emitting diodes,[1] biological markers,[2] and solar cells.[3] Experiments on ZnS quantum dots have looked at the their efficiency as photocatalysts,[4] especially for reactions involving water pollutants.[5–7] Semiconductor quantum dots might even appear in the phosphors of nonmercury based fluorescent lights. The excitation properties of semiconductor quantum dots differ from both small molecules and bulk semiconductor materials; notably, electron energy level spacing varies with quantum dot size.

In addition to size dependent energy spacings, semiconductor quantum dots have electron orbital patterns not found in two or three atom systems, nor do their wave functions have the full symmetry of the bulk material. Effective mass,[8–13] tight binding,[14–20] quasiparticle,[21] and empirical pseudopotential[22–24] models of semiconductor quantum dots provide ways to find patterns relating optical transitions to wave-function shape and size.

We have previously done calculations with a phenomenological empirical pseudopotential model which described band gaps vs size for CdS and CdSe quantum dots with significantly better agreement to experiments than effective mass models for the 10–30 Å radius range.[22,25,26] That pseudopotential model did not include the mixing of bulk wave vector states or detailed surface effects in its Hamiltonian. In the present paper, we employ a real-space representation of the empirical pseudopotential which allows such effects to be addressed. The final Hamiltonian is similar to the pseudopotential Hamiltonian of Wang and Zunger,[27,28] particularly the surface passivation;[28] yet, its semiconductor parameters correspond directly to our previous work[26] and a slight modification of the original empirical pseudopotential form factors[29,30] for ZnS and CdSe. The methodology is validated via calculation of band gaps vs radius for wurtzite ZnS and CdSe quantum dots, and we compare the model's energy shifts to experiments, our previous model, and the model of Wang and Zunger.

In order to examine connections between the optical transitions and quantum dot wave functions, we used two powerful tools, the recursive residue generation method[31] and three dimensional visualization software. The recursive residue generation method (RRGM) can produce squared transition matrix elements from a single electron energy eigenstate to all other eigenstates without the computational cost of solving for the complete eigenvectors of the other states. The RRGM works directly with our use of the Lanczos algorithm on the exponentially transformed Hamiltonian[32] for calculating the energy eigenvalues of the Hamiltonian. Electron–hole interactions can be included when squared matrix elements are computed, and the RRGM can also assist in finding surface states.

The wave functions of eigenstates with specific properties, such as a strong oscillator strength to another state, can be analyzed and sorted with the help of probability density isosurfaces generated by three dimensional data visualization software.

Since spin–orbit effects are currently ignored by our model and the spin–orbit splitting is smaller for ZnS than for CdSe, we focus the analysis of wave functions and squared transition matrix elements on our calculations for wurtzite

ZnS quantum dots; nevertheless, we make some comparisons with the literature on wurtzite CdSe dots on the basis that wurtzite ZnS and wurtzite CdSe have similar crystal lattice structures and valence configurations. We obtain an increase of the integrated oscillator strength of the lowest absorption peak with dot size, qualitatively resembling the experiments of Bawendi and co-workers.[33]

The organization of the paper is as follows: Section II presents our method of calculation. In Sec. III, we compare our lowest exciton calculations with recent experiments and other pseudopotential models. Section IV presents our transition strength calculations for ZnS quantum dots, and Sec. V presents a summary of our conclusions.

## II. THEORETICAL MODEL AND METHOD OF CALCULATION

For this paper, the description of electrons in quantum dots begins with an effective one electron (or hole) Hamiltonian. We calculate the energy eigenvalues of our Hamiltonian using an exponential spectral transform and the Lanczos algorithm.[32] Transforming a Hamiltonian into an exponential form has been shown to speed up convergence of the lowest energy states as compared to application of the simple Lanczos algorithm to an untransformed Hamiltonian.[32] Feit et al.[34] introduced the following spectral operator:

$$\mathbf{A}_H = e^{(-\beta \mathbf{E}_{\text{potential}}/2)} e^{(-\beta \mathbf{E}_{\text{kinetic}})} e^{(-\beta \mathbf{E}_{\text{potential}}/2)}. \quad (1)$$

For small $\beta$, this operator has eigenvalues that are numerically close to the exact exponential $e^{-\beta \mathbf{H}}$. The exact exponential operator $e^{-\beta \mathbf{H}}$ has the same eigenvectors as the Hamiltonian $\mathbf{H}$; hence, one can take the logarithm of $\mathbf{A}_H$'s eigenvalues to obtain numerically accurate eigenvalues of $\mathbf{H}$. For the Lanczos algorithm with no reorthogonalization, we use the Cullum and Willoughby spurious test[35] to eliminate false eigenvalues.

The wave function basis is a real-space grid of $64 \times 64 \times 64$ points for our smaller quantum dots and $128 \times 128 \times 128$ points for our larger structures. In applying the Hamiltonian for the Lanczos recursions, Fast Fourier transforms convert the grid into momentum space, where the kinetic energy operator is diagonal. After the potential was generated for a 204 atom ZnS cluster, the 1600 Lanczos recursions to converge the two band edge states to within 1 meV took 537.6 seconds on a single 256 Mb IBM SP2 processor. This time would be substantially smaller for a more modern processor.

We computed transition amplitudes between the valence band states and the entire conduction band using the recursive residue generation method (RRGM).[31] The RRGM relies on the fact that the first component of an eigenvector for the Lanczos generated tridiagonal matrix corresponds to the vector dot product of that eigenvector with the initial Lanczos iteration starting vector. By setting the initial Lanczos vector to be the result of applying a dipole potential operator on the initial wave function, the first components of the corresponding tridiagonal matrix eigenvectors become proportional to the dipole matrix element with the initial wave function. As suggested in Wyatt and Scott,[31] we used a modified QL algorithm to obtain the first row eigenvector elements as it solves the tridiagonal Lanczos matrix. We have summed the squares of the first components over the multiple Lanczos copies of the states because the Lanczos with QL tridiagonalization procedure distributes the first component amplitude among the multiple copies.[31]

For summations over states in transition amplitude calculations, energy gaps smaller than $10^{-6}$ eV were treated as degeneracies where oscillator strengths add.

We attempted to create a real-space pseudopotential Hamiltonian that carried over the method of RamaKrishna and Friesner[22,25,26] as much as possible, including the use of the same wurtzite CdSe pseudopotential form factors used in our earlier model of CdSe quantum dot energy gaps.[26] Those CdSe pseudopotential form factors and the form factors for wurtzite ZnS used here are from Refs. 29 and 30, with some small changes to adjust the top bulk valence bands. Two nonzero form factors of each material were modified to bring the highest bulk valence band doublet above the highest bulk valence band singlet. The first quantum dot Hamiltonian that we tried in real-space consisted of the reciprocal lattice vector expansion for the empirical pseudopotential inside the quantum dot and a large constant spherical well at the quantum dot surface; however, the boundary conditions with this Hamiltonian produced too many surface states in the band gap for usable results.

We needed more realistic boundary conditions for our quantum dot surface atoms that were more like the decay of a finite atom's potential than the plane wave pseudopotential/spherical well potential interface. To create a decaying potential, we added new Fourier components to the potential by fitting the original empirical pseudopotential nonzero and zero form factors to a sum of Gaussians. Wang and Zunger[27] used a mostly Gaussian form to fit LDA form factors from a number of different crystal structures and then adjusted the potential again to fit the band structure; while our approach was to recreate the original empirical pseudopotential form factors in the bulk limit. The centers and widths of the Gaussians were allowed to vary in our fitting procedure, unlike the fixed centers and widths used in Ref. 27. Extra points from splines and/or point weighting were sometimes used to reduce the appearance of Gaussians with very narrow widths (having a range much larger than the bond length in real-space) and sets of large amplitude Gaussians that cancel each other in the pseudopotential. The long wavelength, $\|\mathbf{G}\|=0$ values for the wurtzite ZnS and CdSe potentials were chosen to place the bulk work functions within 0.5 eV of the experimental values[27] reported in Ref. 36 for wurtzite CdSe and zinc-blende ZnS. The pseudopotential work functions are $-7.115$ eV and $-7.500$ eV for CdSe and ZnS, respectively. We do not expect 1 eV errors in the bulk work function to significantly affect quantum dot interior electronic states. Our form factors are listed in the appendix. The bulk band structure from our Gaussian fitted potentials matches the band structure from our original form factors to within 0.003 eV along the $\Gamma$ to $\mathbf{A}$ and $\Gamma$ to $\mathbf{M}$ lines of the 16 lowest energy bands (475 $\mathbf{G}$ vectors were used). For the quantum dot calculations, the pseudopotentials were Fourier transformed to yield radial potentials for each atom type:

$$V_{\text{Cd,Zn}}(\|\mathbf{R}\|) = 4\pi K \int_0^\infty g^2 (V_\mathbf{S}(g) + V_\mathbf{A}(g)) \frac{\sin(g\|\mathbf{R}\|)}{g\|\mathbf{R}\|} dg,$$

$$V_{\text{Se,S}}(\|\mathbf{R}\|) = 4\pi K \int_0^\infty g^2 (V_\mathbf{S}(g) - V_\mathbf{A}(g)) \frac{\sin(g\|\mathbf{R}\|)}{g\|\mathbf{R}\|} dg, \quad (2)$$

where

$$K = \frac{\Omega(\text{unit cell volume})}{n(2\pi)^3}.$$

The potential due to each atom was interpolated from a stored set of radial values in order to reduce computation time. The quantum dot Hamiltonian with these atom centered potentials does not need a spherical confining potential.

Our CdSe pseudopotential gives a bulk band gap of 1.883 eV. The experimental bulk gap is 1.829 eV at 293 K and 1.751 eV at 80 K.[37] For dot calculations we have shifted our lowest exciton transition energies by a small constant to achieve a bulk limit of 1.79 eV, the average of 1.829 eV and 1.751 eV. The ZnS pseudopotential gives a bulk gap of 3.682 eV. The $\Gamma_{5\,\text{valence}}$ to $\Gamma_{3\,\text{conduction}}$ transition energies are 4.226 eV and 5.696 eV for CdSe and ZnS, respectively.

The ''spherical'' quantum dots under study have the unit cell arranged such that the atom type exchanges upon inversion about the origin. The unique wurtzite $c$ axis is along the $\hat{z}$ direction for all quantum dots in this work. In building the quantum dot, atoms were generated within a sphere, and some surface atoms were removed. For all the quantum dots in this work, we defined the radius as the average over all semiconductor atoms with dangling bonds of the atom centered radii from the origin.

Our surface passivation potentials are based upon Refs. 28 and 38. The ZnS and CdSe passivation potentials are identical, but the distances between the surface semiconductor atoms and the passivation potential centers were scaled for ZnS by the ZnS:CdSe bond length ratio. Surface atoms with three dangling bonds were removed because we found that band edge valence states of CdSe quantum dots were distributed more deeply inside the dot than without their removal. A more systematic approach that determines the passivation potentials for arbitrary empirical pseudopotential systems would be desirable in future work with this type of model.

For the exciton energy calculations in the next section, in order to take into account the Coulomb attraction of an excited electron–hole pair, the term

$$\frac{-1.786}{\epsilon R} \text{hartrees bohr} \quad (3)$$

was added to the single particle quantum dot band gap from the empirical pseudopotential.[9] $\epsilon$ is the high frequency dielectric constant. For hexagonal CdSe we set $\epsilon$ to 6.25, the average of $\epsilon_\parallel$ and $\epsilon_\perp$.[37] For hexagonal ZnS, we set $\epsilon$ to 5.415, the average of the two values from the compilation edited by Madelung.[39]

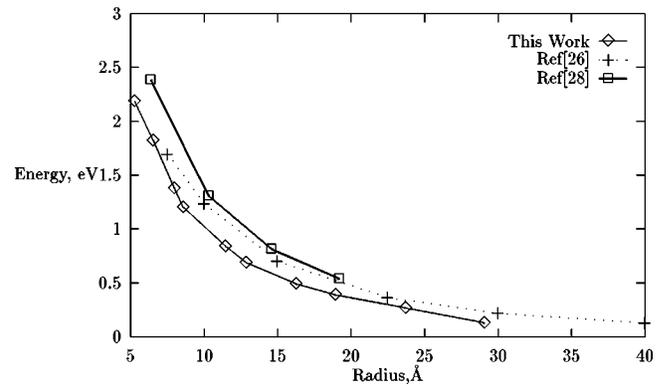

FIG. 1. Wurtzite CdSe single particle gap shift from the bulk value for three models. Reference 26 is a model that uses empirical pseudopotential bulk energy dispersion to approximate quantum dot gaps. Reference 28 is another empirical pseudopotential model. The electron–hole Coulomb term is not included here.

### III. LOWEST EXCITON ENERGY CALCULATIONS FOR CdSe AND ZnS DOTS

Figure 1 shows the shift in the single particle band gap from the bulk value in comparison to the pseudopotential model of Ref. 28 and our previous model.[26] The difference in band gap shift between this work and our previous model is smaller than the difference between the model and the pseudopotential model of Wang et al.[28] The dominant contribution to the spectral shift for CdSe and ZnS quantum dots comes from the conduction band.

In Fig. 2 we plot the model's lowest exciton energies for our CdSe quantum dots and results from experiments with narrow size distributions.[43,54,55] The agreement is good with the optical absorption experiments, but not as good with the single dot conductance tunneling spectroscopy results of Alperson et al. Optical experiments with single quantum dots may help to resolve the disagreement between the experiments.

For ZnS clusters, the computed single particle band gaps are 5.90, 5.53, 4.91, 4.57, 4.39, 4.19, 4.08, 3.93, 3.88 eV for radii of 5.01, 6.56, 8.92, 11.43, 13.74, 16.65, 19.15, 24.10, 27.50 Å, respectively.

Figure 3(a) shows our ZnS quantum dot exciton energies, with the electron–hole correction of formula (3), as compared to experiments reporting the wurtzite crystal structure.[5,6,40,56,57] As with CdSe quantum dots, the model describes an increase in absorption energy with decreasing quantum dot size for the radii considered. Experimental data in Ref. 40 show a gap increase with decreasing size in the region near 4 eV. Significantly more experiments involved zinc-blende crystal structure ZnS quantum dots. A different pseudopotential is required for the zinc-blende crystal structure. Our calculations in a previous paper[26] gave similar lowest exciton energy shifts from the bulk for wurtzite and zinc-blende CdSe quantum dots. We have included a comparison of our wurtzite model with zinc-blende ZnS quantum dot experiments[53,58–64] in Fig. 3(b). The result of a nearest-neighbor tight-binding calculation[14] is lower than our model and most of the experiments.

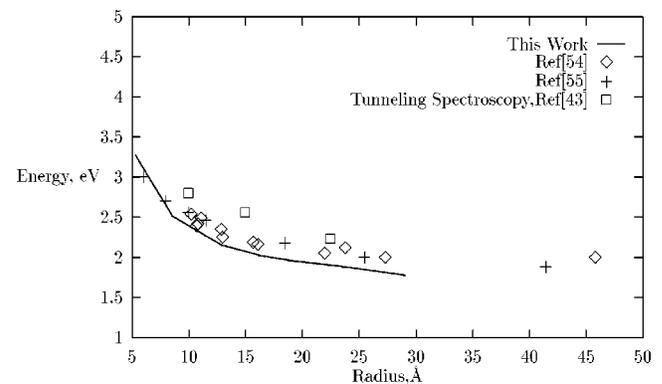

FIG. 2. Theoretical lowest exciton energies vs experiments on wurtzite CdSe quantum dots.

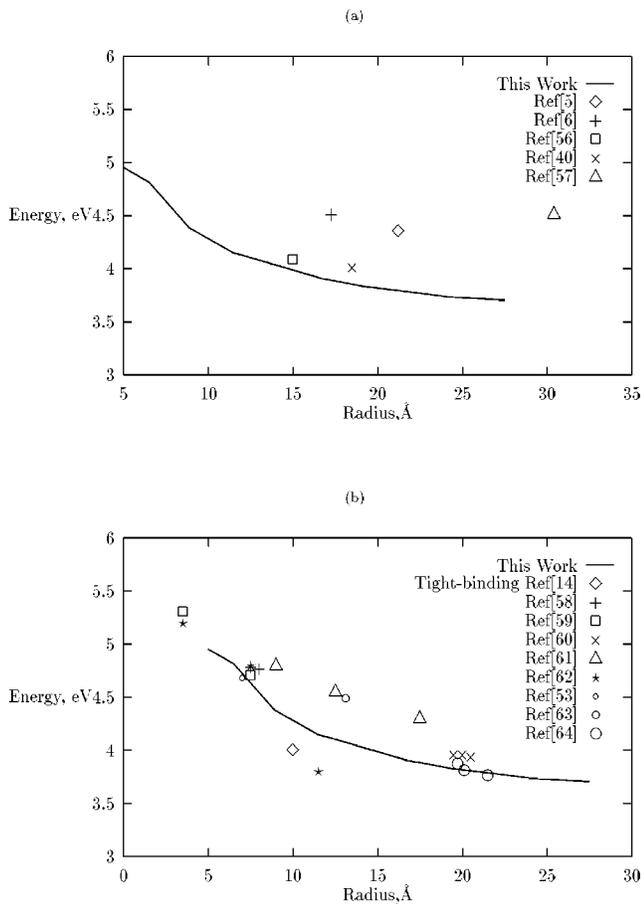

FIG. 3. (a) Theoretical lowest exciton energies vs experiments on wurtzite ZnS quantum dots. The reported energy of Ref. 40 is between 3.97 and 4.05 eV. (b) Theoretical lowest exciton energies vs experiments on zinc-blende quantum dots. Reference 53 could not report a definitive determination of crystal structure through its x-ray analysis.

## IV. TRANSITION STRENGTH AND WAVE-FUNCTION SYMMETRY IN ZnS DOTS

In the dipole approximation, the electron transition oscillator strength is proportional the transition energy and $\|\langle \Psi_{\text{valence}} | \vec{r} | \Psi_{\text{conduction}} \rangle\|^2$,[41] the squared transition dipole sum. We have looked for patterns in the wave functions of ZnS quantum dots with various radii and tried to correlate wave function symmetry with transition strength.

The charge density isosurface labeled $|\Psi_a\rangle$ in Fig. 4 depicts a state near the valence band edge that consists of "$p$" like orbitals oriented mostly parallel to a line 30 degrees from the $y$ axis. The vector defined by

$$\sum_{\vec{r}} x \Psi(\vec{r}) \hat{\mathbf{x}} + y \Psi(\vec{r}) \hat{\mathbf{y}} \qquad (4)$$

has the direction shown by the arrow inside the isosurface of the state $|\Psi_a\rangle$. This type of state and the state oriented perpendicular to it in the $x$–$y$ plane are labeled the $Y'$ and $X'$ linear states in this paper. In the bulk semiconductor, these two states are degenerate, and the corresponding $Z$ axis state is split off by the unequal crystal field. We found the $Y'$ and $X'$ linear states, with some mixing as described later, at the top of the valence band for 84 to 504 atom ZnS quantum dots. In dots that are elongated along the distinct $z$ axis, the $Z$ linear state can be expected to move to the top of the valence band and above the $X'$ and $Y'$ linear states because $\hat{z}$ would be a less confined direction. Pseudopotential calculations for ellipsoidal CdSe quantum dots have shown this ordering change with elongation of the $z$ axis.[42]

The splitting of the $X'$ and $Y'$ linear states from the bulk degeneracy seems plausible because our quantum dots are not perfectly symmetric in the $x$–$y$ plane. Simple effective mass models may miss this type of splitting. In contrast to our results, tight binding calculations reported the states as degenerate for surface relaxed wurtzite CdSe quantum dots.[20]

The other type of valence state that we considered in detail consists of "$p$" like orbitals rotating about the $z$ axis, and this $Z$ angular state is depicted in Fig. 4(b). It is the highest valence state for the 46 atom quantum dot and the third or fourth highest, including mixtures with other states, for the 84 atom to 484 atom quantum dots. Angular states in other orientations and with nodes at various positions were found in all of the quantum dots. Due to the antisymmetry of the $Z$ angular states, their lowest energy strong transitions to the conduction band are to conduction band states with $P$ symmetry envelope functions in the $x$–$y$ plane. Like the $X'$ and $Y'$ linear valence states, the $P_{x'}$ and $P_{y'}$ conduction states are nondegenerate and are oriented at an angle from our coordinate axes. Such a splitting in the $P_{x'}$ and $P_{y'}$ would change the peak spacing in tunneling current-voltage spectroscopy experiments.[43] (A corresponding $P_z$ conduction band state, split off by the crystal field and quantum dot shape can be seen in Fig. 8.) The dipole transitions from the $Z$ angular valence state to the $P_{x'}$ and $P_{y'}$ states have maximum amplitude for light linearly polarized parallel to both the plane that divides the $P$ envelope lobes of the unoccupied state and the $x$–$y$ plane. Circularly polarized light would excite the $Z$ angular valence electron into both $P_{x'}$ and $P_{y'}$ conduction states, producing two absorption peaks of similar magnitude.

The squared $x$–$y$ circular polarization matrix elements, normalized per semiconductor atom, of transitions from the $X'$ and $Y'$ linear valence states to the lowest conduction state, which has $S$ like envelope function symmetry, were plotted as a function quantum dot size in Fig. 5(a). The transitions are polarization dependent, with a maximum ampli-tude for light polarized close to the direction of the arrow defined by formula (4). Figure 5(a) shows that for both type of wave functions, the overall squared dipole sum per atom decreases with quantum dot size.

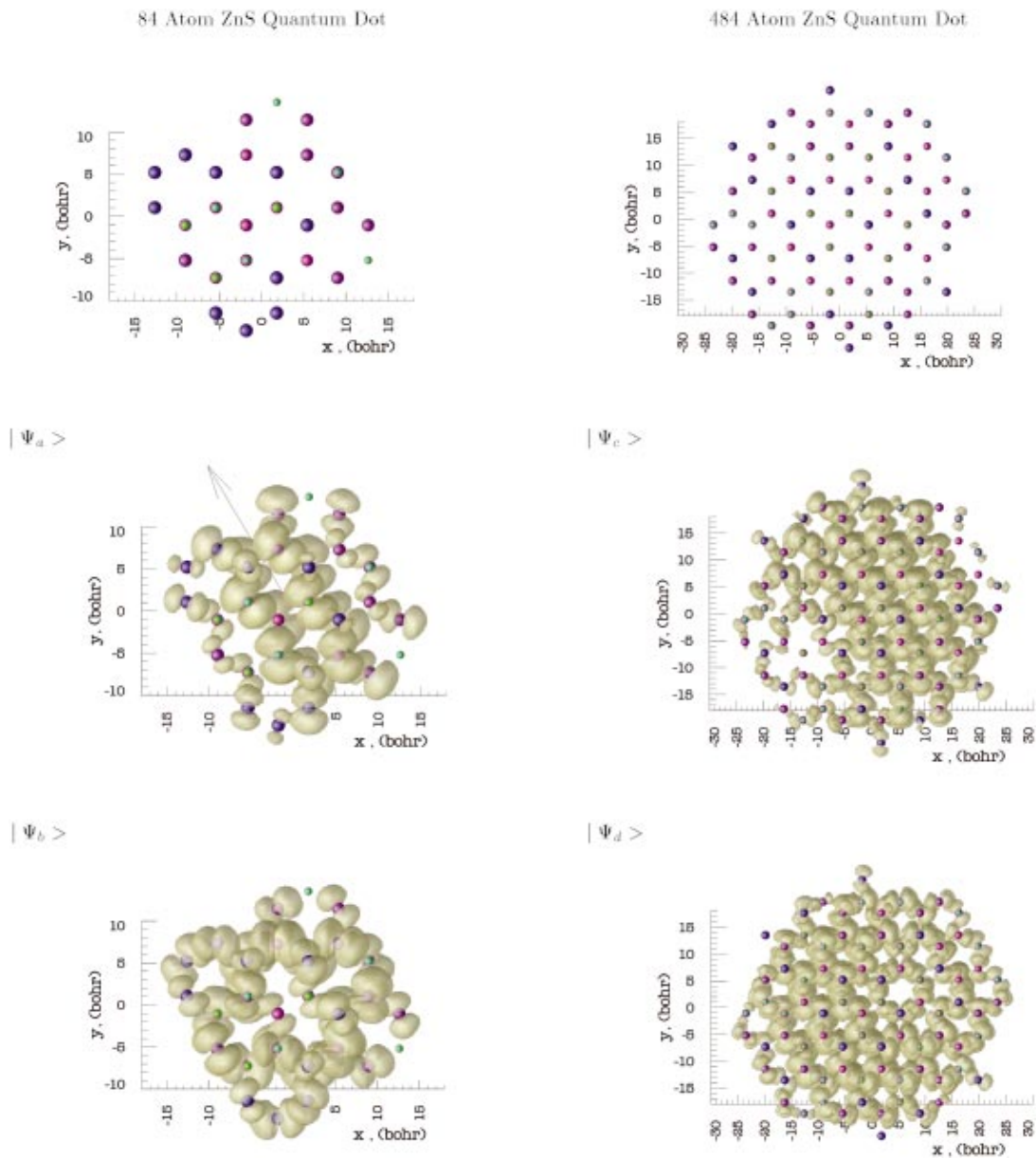

FIG. 4. (Color) Two different ZnS quantum dots and corresponding valence states. Red/violet spheres represent S atoms, and green spheres represent Zn atoms. The gray electron charge density isosurfaces $|\Psi a\rangle$ and $|\Psi b\rangle$ represent the $Y'$ linear and $Z$ angular valence states of the 84 atom quantum dot on the top left. The isosurfaces $|\Psi c\rangle$ and $|\Psi d\rangle$ represent the mixed $Y'$ linear and $Z$ angular valence states of the 484 atom quantum dot on the top right.

How does the electron–hole Coulomb interaction affect the matrix elements? In order to include the interaction of the excited electron with the hole that starts from our initial state vacancy, we have developed an approximate electron–hole interaction potential. We used a multipole based electrostatic potential far away from each atom and a direct grid sum for points close to an atom. The multipoles have a lower computational cost than double summation over the real-space grid. A spherical region of grid points centered about each atom is used to sum the multipole moments. For the highest valence band states, we found that terms up to the quadrapole describe the double lobed $p$-orbitals, shown in Fig. 4, quite adequately. In terms of the nearest-neighbor distance $\delta$, the radii of the spherical regions for multipole summation were $0.85\delta$ for sulphur and $0.15\delta$ for zinc. There is a small amount of both double counting and missed charge density in this scheme. For the valence edge states considered in this work, only a small fraction of the total charge is excluded by ignoring the points between the spherical regions. Multipole terms are used for distances greater than $1.1\delta$, which provides a buffer zone where the direct grid sum from each atomic sphere crosses over to the multipole region. All potential terms are divided by the high frequency dielectric constant. It would be desirable to include the fully position dependent dielectric screening with effects of the nanostructure boundary in future calculations. Outside the quantum dot (beyond a distance of the largest radius of the semiconductor atoms from the origin $+1.1\delta$), the electron–hole po-tential is damped by multiplication with an unnormalized Gaussian with $\sqrt{2}\sigma = 0.05\delta$.

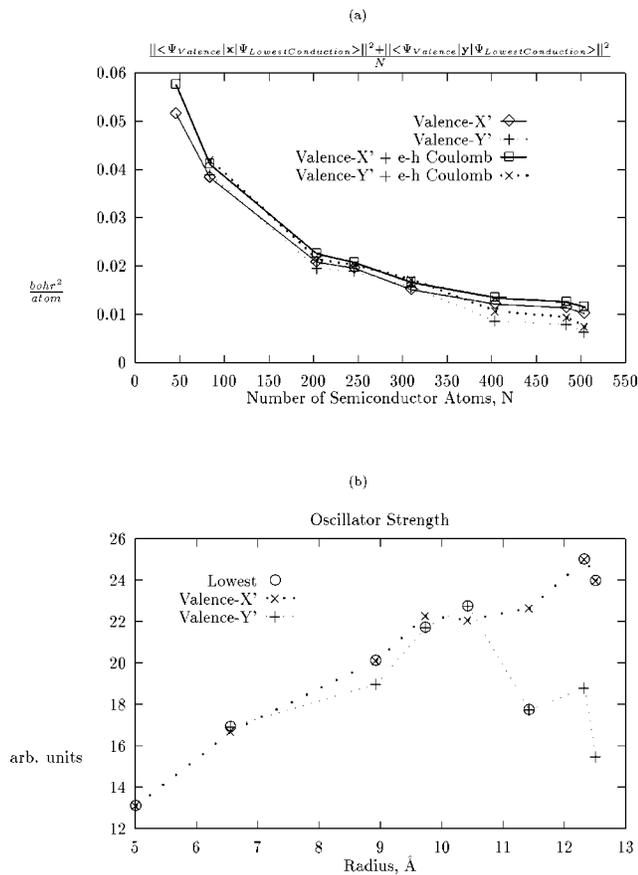

FIG. 5. (a) Comparison of matrix elements squared, normalized per semiconductor atom, for linear dipole valence states to the lowest conduction band, with and without $e-h$ Coulomb effects. (b) Comparison of the total oscillator strength vs ZnS quantum dot radius for the same transitions with the lower energy transition denoted by circles.

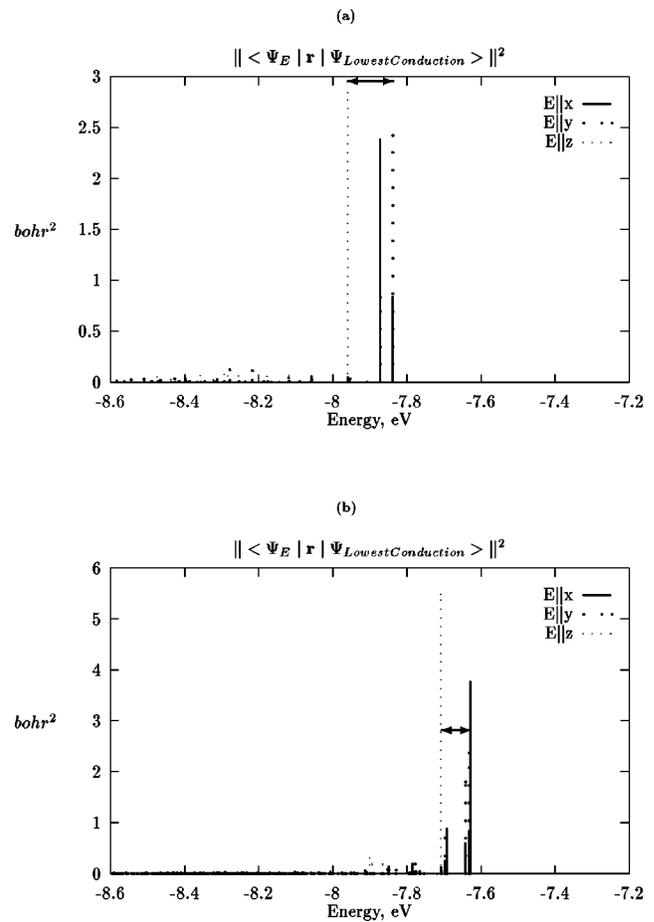

FIG. 6. The squared $x$, $y$, and $z$ polarization matrix elements between the lowest unoccupied state and the valence states as function of valence state energy. (a) and (b) are for 6.56 Å and 12.51 Å radii quantum dots, respectively.

The electron–hole Coulomb interaction shifts the conduction band states to lower energies. The inclusion of the effective electron–hole interaction increased the electron–hole overlaps and the squared matrix elements in Fig. 5(a); however, the overall scaling with dot size did not change significantly.

In contrast to the small difference in per atom scaling between the two transitions of Fig. 5(a), Fig. 5(b) shows a significant difference between the total oscillator strength vs size trends of the two transitions. The oscillator strengths of Fig. 5(b) use energies with the perturbative Coulomb term and squared matrix elements which include the effects of the approximate electron–hole Coulomb interaction. For the linear $X'$ state, the squared matrix element appears to increase over most of the size range, while the $Y'$ state squared matrix element increases with radius most of the way to 10.42 Å and decreases at larger sizes. The disparity between the two linear states occurs primarily from the mixing of the $Y'$ linear state with the $Z$ angular state in the 300–500 atom size range. Electron density isosurfaces of the $Y'$ state mixed with the $Z$ angular state for the 484 atom dot, and the complementary $Z$ angular state below it in energy are shown in $|\Psi_c\rangle$ and $|\Psi_d\rangle$ of Fig. 4. Those mixed even and odd states lead to optical absorption that is a mixture of the pure $Y'$ linear transitions and the $Z$ angular state transitions described earlier.

Recent experimental absorption measurements of wurtzite CdSe quantum dots show an approximately linear increase with radius in the integrated oscillator strength of the lowest ($1S$ valence to $1S$ conduction) absorption peak for radii between 11.8 and 33.9 Å.[33] One simple model predicted that at small radii, the oscillator strength of the lowest transition of a semiconductor quantum dot should not vary much with radius.[12] For a large ensemble of quantum dots with random orientations and broadening due to size variation and phonons, the lowest experimental absorption peak would include excitations from the highest $X'$, $Y'$, and $Z$ linear valence states if the energy splittings among the states are small. For the ZnS quantum dots that we considered, those splittings are tens of meV or smaller. In order to compare our model with future lowest absorption peak transition measurements on ensembles of wurtzite ZnS quantum dots, we computed squared transition dipole sums between the bottom conduction state and the valence band by applying the RRGM on the bottom conduction state. Figures 6(a) and 6(b) show the squared matrix element sums as a function of valence band energy for 84 atom and 504 atom quantum dots.

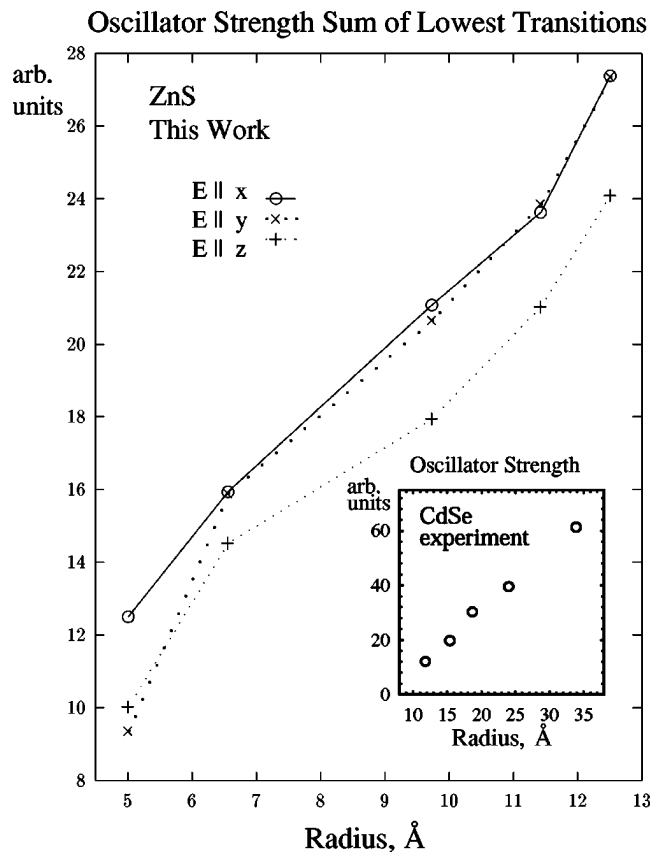

FIG. 7. The total oscillator strength for $x$, $y$, and $z$ polarizations of transitions to the conduction band bottom from valence band states between and including the $Z$ valence state and the top valence state. The inset shows the experimental oscillator strengths of the lowest absorption peak region for ensembles of wurtzite CdSe quantum dots as reported in Ref. 33.

The dipole operator's polarization correlates the highest amplitudes shown with the highest $X'$, $Y'$, and $Z$ linear valence states; and the mixing of the $Y'$ linear valence state with other states at the 504 atom size produces multiple peaks in the $y$ polarization. Summing the oscillator strengths (without electron–hole effects on the matrix elements and using formula (3) for the Coulomb energies) of transitions with initial valence state energies ranging from the $Z$ linear valence state to the top valence state yields the integrated oscillator strength of the lowest absorption peak region for Fig. 7. The double horizontal arrows in Figs. 6(a) and 6(b) represent the valence band summation region. For all three electric field polarizations, our model predicts an increasing integrated oscillator strength of the lowest absorption peak region with increasing radius for ZnS quantum dots with radii between 5 and 12.5 Å. The increase occurs despite the complex mixing of wave-function symmetries discussed earlier. While we have not done the corresponding oscillator strength calculations for CdSe quantum dots because our model lacks spin–orbit coupling, the scaling shown in Fig. 7 resembles the experimentally reported trend of integrated oscillator strength for the lowest absorption peak of CdSe quantum dot ensembles.[33] Experimental measurements of integrated oscillator strength for the lowest transition peak of ZnS quantum dot ensembles can be another future test of the pseudopotential model.

In addition to mixing of even and odd angular states, a lack of symmetry in the electron charge distribution due to the polar crystallite structure of wurtzite quantum dots also appeared in some of our ZnS quantum dot wave functions. The polar asymmetry can allow dipole transitions that would otherwise be forbidden by dipole selection rules. For example, pure $X'$ and $Y'$ linear valence states would have zero dipole transition strength to the conduction states with $P_z$ like envelope functions. In other words, the dipole sums for different regions of the quantum dot cancel each other out. Our ZnS quantum dots have Zn on the top layer for positive $z$ and S on the bottom layer for negative $z$, and there is net $z$ displacement between Zn and S layers throughout the quantum dots. For many sizes of ZnS quantum dots, we found that the charge distribution for the $X'$ and $Y'$ linear states was unevenly biased towards negative $z$ values. Figure 8 shows an electron density isosurface of the $Y'$ linear valence state for a 246 atom, 9.74 Å radius quantum dot, looking along the $z$ axis. A charge density isosurface of the lowest energy conduction $P_z$ like envelope function state is also shown. The arrows in Fig. 8 illustrate that the upper hemisphere contribution to the dipole matrix element between the two states shown cancels only 51% of the bottom hemispheres matrix element contribution. For this quantum dot size, the lowest energy $P_z$ conduction state's energy lies between the conduction bottom state and the split lowest $P_{x'}$ and $P_{y'}$ conduction states. Using our approximate electron–hole Coulomb interaction, the energies of the $Y'$ linear valence to conduction bottom transition and the corresponding transition to the $P_z$ state are 4.48 eV and 5.07 eV, respectively. These energies and our computed matrix elements for circular polarization in the $x$–$y$ plane with the electron–hole Coulomb interaction suggest that the transition to the $P_z$ state has 2.5% of the oscillator strength of the transition to the bottom conduction state.

A number of experiments for wurtzite CdSe quantum dots have looked at infrared transitions believed to be from the lowest $S$ envelope conduction band state to the lowest $P$ envelope conduction states.[44–47] By the rules that define matrix element multiplication, the broken symmetry allowed $X'$ or $Y'$ valence to $P_z$ transitions discussed above for wurtzite ZnS quantum dots imply that the $S$ to $P_z$ conduction intraband transition would be allowed for polarizations in the $x$–$y$ plane, albeit weaker than the $S$ to $P_{x'}$ or $P_{y'}$ transitions with the same polarization.

As shown in our examples and in other work[16,48–50] the transitions for semiconductor quantum dots display complex patterns.

### A. Transitions to conduction band surface states

One way to study excitations to surface states involves the removal of surface passivation from quantum dot atoms. Dipole transitions to a band gap surface state below the conduction bottom state have been studied for a pseudopotential model of cubic InP quantum dots in this way.[51] In that study, the dipole transition strength from the InP quantum dot valence edge state to the In dangling bond surface state decreased with quantum dot size.[51]

Given the existence of a surface state in the quantum dot excitation spectrum, what factors other than quantum dot size affect the transition amplitude to the state? Wave functions that have nodes in the center of a quantum dot, such as the wave functions with angular momentum about the center, might have high overlaps with surface states and thus have high transition strengths to those states. For our investigation of this idea, we studied 84 and 246 atom quantum dots that have two zinc atoms with two surface dangling bonds in their middle $z$-axis layer. For each case, we removed the passivation potentials of one zinc atom at a time. In order to identify surface states in the conduction band, the overlaps with an initial Lanczos vector centered on the unpassivated atom were computed by the recursive residue generation method and plotted. We picked the highest overlap peak in the conduction band as the surface state to study for each quantum dot size and passivation. Table I shows the squared dipole sums from three different valence states to the selected surface states.

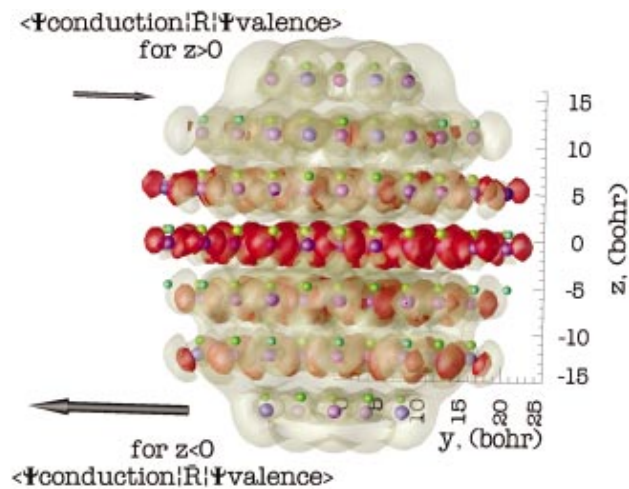

FIG. 8. (Color) The red charge density isosurface is for the top $Y'$ linear valence state of a 246 atom ZnS quantum dot. The gray isosurface corresponds to the $P_z$ conduction band state. The vectors on the left represent transition matrix element sums of the upper and lower hemispheres of the quantum dot.

TABLE I. The squared $x-y$ circular polarization dipole matrix element of transitions to surface states from three different valence states.

| Size | Type of valence $\Psi$ | Surface state/Passivation label | (bohr)$^2$ |
| --- | --- | --- | --- |
| 84 atoms | $Y'$ linear | $a$ | 0.135 |
|  | $X'$ linear |  | 0.035 |
|  | $Z$ angular |  | 0.014 |
|  | $Y'$ linear | $b$ | 0.129 |
|  | $X'$ linear |  | 0.036 |
|  | $Z$ angular |  | 0.014 |
| 246 atoms | $Y'$ linear | $c$ | 0.005 |
|  | $X'$ linear |  | 0.014 |
|  | $Z$ angular |  | 0.004 |
|  | $Y'$ linear | $d$ | 0.010 |
|  | $X'$ linear |  | 0.009 |
|  | $Z$ angular |  | 0.004 |

Surface states $a$ and $c$ correspond to the removal of two repulsive passivation potentials for a zinc atom in the middle $z$-axis layer and positive $\hat{y}$ surface of the quantum dot. Surface states $b$ and $d$ correspond to middle layer surface atoms which are $-67.6$ and $-54.8$ degrees from the $x$-axis, respectively. Ignoring the Coulomb shift, the surface states $a$ and $b$ from the 84 atom structure were both 6.71 eV above the valence edge (1.19 eV above the conduction edge). States $c$ and $d$ of the 246 atom dot were both 6.98 eV above the valence edge. In Table I, the angular state clearly has a lower transition probability to the surface states studied than the two linear states. For the core dipole transitions described in Fig. 5, the removal of passivation from a two dangling bond zinc atom changed the squared matrix elements by less than 5%.

## V. CONCLUSION

We have presented energy gap calculations for wurtzite CdSe quantum dots with up to 4096 atoms and for wurtzite ZnS quantum dots with up to 4898 atoms. We successfully used real-space potentials based upon empirical pseudopotential form factors to carry out these calculations.

The calculated lowest transition energies for CdSe quantum dots were close to the energies from our single band empirical pseudopotential approach, the pseudopotential calculations of Wang et al.,[28] and the optical experiments. The conductance tunneling spectroscopy energies were a fraction of an eV higher than the optical experiments and our model for reasons we do not understand at this time. Future experiments with highly uniform ZnS quantum dot samples may determine the accuracy of the ZnS band gap model more definitively than the comparisons presented here which show good agreement with some optical experiments and poor agreement with others.

Our ZnS quantum dots showed small energy splittings due to broken symmetry in the plane perpendicular to the unique wurtzite $c$ axis. The splitting of states between $X'$ and $Y'$ directions might disappear or change significantly due to surface relaxation and changes in geometry; thus it may be the case that for many of the presently available quantum dot samples for experiments, there is enough symmetry to keep the $X'$ and $Y'$ states mixed and degenerate. Nevertheless, slight asymmetries in the $x-y$ plane, perpendicular to the unique bulk axis, can split the degeneracy in that plane as our model for ZnS quantum dots describes.

For ZnS quantum dots, we computed transition matrix elements using the recursive residue generation method.

Both the mixing of even and odd angular states and charge asymmetry along the unique wurtzite axis affected interband and some intraband transitions in complex and polarization dependent ways. Despite the changes in individual state transitions caused by those factors, the total oscillator strength of the lowest energy peak region increased with radius at small radii.

Finally, matrix element calculations showed that excitations from angular valence band states to conduction band surface states associated with unpassivating surface bonds do not necessarily have higher transition rates than excitations from valence states without central nodes.

Of course, one of the main limitations of the current model is that it does not quantitatively describe the exact energies and wave functions of the quantum dot surface states. Mixed *ab initio*/empirical or pure *ab initio* calculations may provide those surface state details in the future.

We also neglected both spin–orbit coupling and the electron–hole exchange interaction, which have been included, sometimes perturbatively, in other wurtzite quantum dot electronic structure calculations.[13,20,48,52] Although we have discussed details related to the fine structure of quantum dot optical absorption, our model cannot quantitatively describe all aspects of exciton fine structure such as a dipole forbidden dark state for the lowest interband transition[13,20,48,52] without adding those factors into our model. The effect of electron–hole exchange on the energy levels for wurtzite CdSe quantum dots has been studied via pseudopotentials with configuration interaction included.[48] Despite the drawbacks of the calculations presented here, they elucidate further questions regarding patterns in the electronic transitions of wurtzite semiconductor quantum dots for theorists and experimentalists to explore.

## ACKNOWLEDGMENT

This work was partially supported by a grant from the Department of Energy (DE-FG02-90ER14162).

## APPENDIX: PSEUDOPOTENTIAL PARAMETERS

Using atomic units, the form factors for wurtzite ZnS (lattice parameter$=3.814$ Å,[30] ideal wurtzite lattice) are

$$V_S(\|\mathbf{G}\|, \text{Bohr}^{-1}) = -0.513\,7349\,e^{(-((\|\mathbf{G}\|-0.502\,7628\times 10^{-2})/0.813\,4720)^2)} + 0.342\,7305\times 10^{-1}\,e^{(-((\|\mathbf{G}\|-2.102\,973)/0.221\,6043)^2)}$$
$$+ 0.305\,3194\times 10^{-1}\,e^{(-((\|\mathbf{G}\|-1.837\,589)/0.145\,1529)^2)} - 0.218\,2857\times 10^{-1}\,e^{(-((\|\mathbf{G}\|-1.125\,889)/0.113\,1840)^2)}$$
$$- 0.113\,6577\times 10^{-1}\,e^{(-((\|\mathbf{G}\|-1.843\,26)/0.077\,035\,58)^2)} - 0.977\,4217\times 10^{-2}\,e^{(-((\|\mathbf{G}\|-1.477\,191)/0.119\,6223)^2)}$$
$$+ 0.195\,0457\times 10^{-2}\,e^{(-((\|\mathbf{G}\|-2.277\,19)/0.082\,228\,76)^2)} - 0.433\,5315$$
$$\times 10^{-3}\,e^{(-((\|\mathbf{G}\|-2.535\,979)/0.084\,948\,76)^2)} \quad \text{hartrees,} \tag{A1}$$

$$V_A(\|\mathbf{G}\|) = 0.238\,8607\,e^{(-((\|\mathbf{G}\|-0.484\,6192)/0.681\,9721)^2)} + 0.227\,9054\times 10^{-1}\,e^{(-((\|\mathbf{G}\|-1.778\,875)/0.119\,1104)^2)}$$
$$+ 0.227\,609\,55\times 10^{-1}\,e^{(-((\|\mathbf{G}\|-2.105333)/0.223\,6762)^2)} + 0.211\,3845\times 10^{-1}\,e^{(-((\|\mathbf{G}\|-1.498\,629)/0.135\,7484)^2)}$$
$$- 0.130\,1762\times 10^{-1}\,e^{(-((\|\mathbf{G}\|-2.088\,037)/0.244\,5309)^2)} - 0.613\,0098\times 10^{-2}\,e^{(-((\|\mathbf{G}\|-2.187\,602)/0.069\,804\,04)^2)}$$
$$+ 0.318\,8180\times 10^{-2}\,e^{(-((\|\mathbf{G}\|-2.408\,863)/0.094\,813\,32)^2)} - 0.291\,6796\times 10^{-3}\,e^{(-((\|\mathbf{G}\|-2.535\,020)/0.068\,080\,70)^2)}$$
$$+ 0.134\,6894\times 10^{-3}\,e^{(-((\|\mathbf{G}\|-2.051\,840)/0.056\,25832)^2)} \quad \text{hartrees.} \tag{A2}$$

The form factors for wurtzite CdSe (lattice parameter$=4.299$ Å,[30] ideal wurtzite lattice) are

$$V_S(\|\mathbf{G}\|) = -0.410\,5160\,e^{(-((\|\mathbf{G}\|-0.599\,2449\times 10^{-1}/0.735\,9207)^2)} - 0.278\,8324\times 10^{-1}\,e^{(-((\|\mathbf{G}\|-1.396\,006)/0.104\,7704)^2)}$$
$$- 0.228\,3883\times 10^{-1}\,e^{(-((\|\mathbf{G}\|-1.001\,152)/0.126\,8664)^2)} + 0.184\,3375\times 10^{-1}\,e^{(-((\|\mathbf{G}\|-1.638\,971)/0.207\,4907)^2)}$$
$$+ 0.160\,2960\times 10^{-1}\,e^{(-((\|\mathbf{G}\|-1.910\,275)/0.190\,6031)^2)} - 0.354\,0961\times 10^{-2}\,e^{(-((\|\mathbf{G}\|-1.650\,597)/0.565\,6434\times 10^{-1})^2)}$$
$$- 0.278\,5034\times 10^{-2}\,e^{(-((\|\mathbf{G}\|-2.106\,821)/0.118\,4589)^2)} \quad \text{hartrees,} \tag{A3}$$

$$V_A(\|\mathbf{G}\|) = 0.204\,7268\,e^{(-((\|\mathbf{G}\|-0.450\,4488)/0.607\,9229)^2)} + 0.229\,0629\times 10^{-1}\,e^{(-((\|\mathbf{G}\|-1.760\,302)/0.333\,2508)^2)}$$
$$+ 0.131\,4814\times 10^{-1}\,e^{(-((\|\mathbf{G}\|-1.310\,460)/0.109\,0151)^2)} - 0.126\,78915\times 10^{-1}\,e^{(-((\|\mathbf{G}\|-1.018\,206)/0.147\,5072)^2)}$$
$$+ 0.560\,3843\times 10^{-2}\,e^{(-((\|\mathbf{G}\|-2.056\,526)/0.113\,2995)^2)} - 0.364\,8845\times 10^{-2}\,e^{(-((\|\mathbf{G}\|-2.131\,166)/0.202\,6998)^2)}$$
$$+ 0.302\,1674\times 10^{-2}\,e^{(-((\|\mathbf{G}\|-1.858\,897)/0.056509\,28)^2)} \quad \text{hartrees.} \tag{A4}$$